\documentclass[pra,amssymb,amsmath,twocolumn,a4paper,superscriptaddress,showpacs]{revtex4}

\usepackage{amsfonts,amssymb,amsmath,bm,graphicx,color,mathptmx,hyperref}

\newcommand{\beq}{\begin{equation}}
\newcommand{\eeq}{\end{equation}}
\newcommand{\beqa}{\begin{eqnarray}}
\newcommand{\eeqa}{\end{eqnarray}}
\newcommand{\ket}[1]{| #1 \rangle}
\newcommand{\bra}[1]{\langle #1 |}

\newcommand{\szero}{\hat{S}_0}
\newcommand{\sx}{\hat{S}_1}
\newcommand{\sy}{\hat{S}_2}
\newcommand{\sz}{\hat{S}_3}
\newcommand{\si}{\hat{S}_j}
\newcommand{\sind}[1]{\hat{S}_{#1}}
\newcommand{\fluct}[1]{\hat{\Delta}_{#1}}
\newcommand{\op}[1]{\hat{#1}}
\newcommand{\Tr}{\mathop{\mathrm{Tr}}\nolimits}
\newcommand{\oprho}{\hat{\varrho}}
\newcommand{\ah}{\hat{a}_H}
\newcommand{\ahd}{\hat{a}_H^\dagger}
\newcommand{\av}{\hat{a}_V}
\newcommand{\avd}{\hat{a}_V^\dagger}
\newcommand{\eigen}{\lambda}
\newcommand{\nx}{n_1}
\newcommand{\ny}{n_2}
\newcommand{\nz}{n_3}
\begin{document}

%\draft

\title{Central-moment description of polarization for quantum states of light}

\author{G. Bj\"{o}rk}
\affiliation{Department of Applied Physics, Royal Institute of Technology (KTH)\\
AlbaNova University Center, SE-106 91 Stockholm, Sweden}

\author{J.~S\"{o}derholm}
\affiliation{Max Planck Institute for the Science of Light,
G\"{u}nther-Scharowsky-Stra{\ss}e 1, Bau 24,
91058 Erlangen, Germany}

\author{Y.-S. Kim}
\affiliation{Pohang University of Science and Technology (POSTECH), Pohang, 790-784, Korea}

\author{Y.-S. Ra}
\affiliation{Pohang University of Science and Technology (POSTECH), Pohang, 790-784, Korea}

\author{H.-T.~Lim}
\affiliation{Pohang University of Science and Technology (POSTECH), Pohang, 790-784, Korea}

\author{C.~Kothe}
\affiliation{Department of Physics, Technical University of Denmark, Building 309, 2800 Lyngby, Denmark}

\author{Y.-H.~Kim}
\affiliation{Pohang University of Science and Technology (POSTECH), Pohang, 790-784, Korea}

\author{L.~L.~S\'{a}nchez-Soto}
\affiliation{Departamento de \'{O}ptica,
Facultad de F\'{\i}sica,
Universidad Complutense, 28040~Madrid, Spain}

\author{A.~B.~Klimov}
\affiliation{Departamento de F\'{\i}sica,
Universidad de Guadalajara,
44420~Guadalajara, Jalisco, Mexico}

\date{\today}

\begin{abstract}
We present a moment expansion method for the systematic characterization of the polarization properties of quantum states of light. Specifically, we link the method to the measurements of the Stokes operator in different directions on the Poincar\'{e} sphere, and provide a method of polarization tomography without resorting to full state tomography. We apply these ideas to the experimental first- and second-order polarization characterization of some two-photon quantum states. In addition, we show that there are classes of states whose polarization characteristics are dominated not by their first-order moments (i.e., the Stokes vector) but by higher-order polarization moments.
\end{abstract}

\date{\today}
%\narrowtext

\maketitle

%\tableofcontents

%%%%%%%%%%%%%%%%%%%%%%%%%%%%%%%%%%%%%%%%%%%%%%%%%%%%%%%%%%%%%%%%%%%%%%%%%%%%

\section{Introduction}

A fundamental property of light is its vector nature. Far from a source, freely propagating light can be approximated by a plane wave, with the electric field directed in the plane perpendicular to the direction of propagation. Already the early pioneers of optics realized that a convenient way of characterizing light is to describe the figure the tip of the electric-field vector traces out in this plane. Stokes established an operational procedure to characterize not only the polarization properties of light, but also to what extent a field is polarized~\cite{Stokes}. The method is still the standard way of assessing polarization, although several generalizations, such as polarization of non-plane~\cite{carozzi,setala,luis 3D,gil} and multi-mode~\cite{karassiov,burlakov,iskhakov} fields, have been developed. A limitation of Stokes' approach is that it only considers the average intensities (or photon numbers) and hence only assesses the first-order polarization moments.

As polarization is a relatively robust degree of freedom, that, moreover, can almost losslessly, cheaply, and easily be transformed, it is very often used for coding and manipulating quantum information. Examples of experiments relying on polarization include quantum key distribution~\cite{Charly,muller}, quantum dense coding~\cite{mattle}, quantum teleportation~\cite{bouwmeester}, quantum
tomography~\cite{francesco}, rotationally invariant states~\cite{radmark}, phase super-resolution~\cite{resch}, and weak measurements~\cite{dixon}. However, many of these experiments use correlation measurements, effectively using second, or higher, polarization moments. Such correlation measurements can give surprising results. For example, states that appear unpolarized (that is, with vanishing Stokes parameters), can show unit visibility polarisation-correlations when rotated on the Poincar\'{e} sphere~\cite{usachev}. Such states have been said to have ``hidden polarization''~\cite{klyshko,klyshko97}. As we shall discuss below, there are actually large classes of such states, and they can be classified by the number of lowest-order moments that are invariant under polarization transformations. We shall refer to such states as $r$th-order unpolarized if the first $r$th moments are all invariant under any polarization rotation.

As hinted by this discussion, the full description of polarization can be sorted into moment orders, and simultaneously (but perhaps less obviously) into excitation manifolds. A convenient and experimentally palatable way to do this is by the use of central moments.

For the three lowest orders, the central moments coincide with the cumulants introduced by Thiele~\cite{Thiele}. Each successive cumulant provides information of statistics not already contained in the lower-order cumulants. They have some advantages over a moment description when making affine transformations, and they also provide a simple method of quantifying the difference between a statistical distribution and its simplest Gaussian approximation~\cite{lauritzen}. (For Gaussian distributions all cumulants of order $\geq 3$ vanish.) Kubo promoted their use in quantum mechanics and thermodynamics~\cite{kubo}, but in polarization optics they have been used rather sparsely~\cite{jaiswal,aoki,Brosseau,cai,cai2}.

Below, we first recall some definitions and notation in Sec.~\ref{Sec: Stokes operators}. In Secs.~\ref{Sec:First order} and \ref{Sec:Second order} we examine how first- and second-order polarization properties can be described in terms of expectation values and central moments, respectively. In the following two sections, \ref{Sec:Third order} and \ref{Sec:Fourth and higher order} we discuss how the formalism can be extended to orders higher than the second. In Sec.~\ref{Sec: Polarization properties} we subsequently discuss the connection between excitation manifolds and polarization data and show that polarization tomography in general requires far less data than full state tomography. We then apply the formalism, both theoretically and experimentally, to certain polarization states in Sec.~\ref{Sec: Menagerie}. In particular, we show that there are many states whose polarization characteristics are dominated not by their first-order moments (i.e., their Stokes vector), but by higher-order polarization moments. For example, for three-photon states there exist six different classes of states with different polarization characteristics. Finally we draw some conclusions from the analysis in Sec.~\ref{Sec:Conclusions}.

\section{Stokes operators and the Stokes vector}
\label{Sec: Stokes operators}

We will build on the classical theory of polarization based on the Stokes parameters. For quantized fields, the Stokes operators~\cite{collett} take the role of the Stokes parameters. They are
\begin{eqnarray}
  \label{Stokop}
  & \szero = \ahd \ah +
  \avd \av \, ,
  \qquad
  \sx = \ah \avd + \ahd \av \, , &
  \nonumber \\
  & & \\
  & \sy = i ( \ah \avd -
  \ahd \av ) \, ,
  \qquad
  \sz = \ahd \ah -
  \avd \av \, , &
  \nonumber
\end{eqnarray}
where $\ah$ and $\av$ are the annihilation operators of
the two orthogonal modes, in the following taken to be linearly horizontally and vertically oscillating electric fields,
respectively. The annihilation operators obey the bosonic
commutation relations
\begin{equation}
 \label{BosonicCommutator}
 [ \hat{a}_j, \hat{a}_k^\dagger ] = \delta_{j k} \, ,
 \qquad j, k \in \{H, V \} \, .
\end{equation}

The average values of the Stokes operators correspond to the Stokes
parameters ($\langle \hat{S}_0 \rangle$, $\langle \hat{\mathbf{S}}
\rangle$), where the Stokes vector operator $\hat{\mathbf{S}}$ is $\hat{\mathbf{S}} =
(\sx, \sy, \sz)$. In terms of the Poincar\'{e} sphere, the definitions (\ref{Stokop}) mean that $\sy$ is the eigenoperator for a circularly polarized field, and thus that the operator is parallel to the axis through the south (left-handed circular) and north pole (right-handed circular) of the sphere. $\sx$ and $\sz$ are the eigenoperators for diagonal and anti-diagonal, and horizontal and vertical, linear polarization, respectively. These operators lie in the equatorial plane of the Poincar\'{e} sphere. The directions of $\sx$, $\sy$, and $\sz$ form a right-handed orthogonal vector set in the Poincar\'{e} space.

The Stokes operators satisfy the
commutation relations of the su(2) algebra:
\begin{equation}
  \label{ccrsu2}
  [ \hat{S}_j , \hat{S}_k ] = i 2 \epsilon_{j k \ell} \hat{S}_\ell \, ,
%  [\sx, \sy] = i 2 \sz \, ,
\end{equation}
%and cyclic permutations.
where $\epsilon_{j k \ell}$ is the Levi-Civita tensor. The non-commuting character of these operators leads to the uncertainty relation \beq 2 \langle \sind{0} \rangle \leq  \langle \hat{\mathbf{S}}^2 \rangle - \langle \hat{\mathbf{S}} \rangle^2 \leq \langle \sind{0} \rangle(\langle \sind{0} \rangle + 2). \label{Eq: Uncert relation}\eeq

In spherical coordinates we can use the polar and azimuthal coordinates $\theta$
and $\phi$ to parameterize the unit vector as $\mathbf{n} = (\sin \theta \cos \phi,\sin \theta
\sin \phi,\cos \theta)$. (Note, however, that $\theta$ is the angle to $\sz$, and that $\sx$ and $\sz$ lie in the equatorial plane of the Poincar\'{e} sphere, as explained above.) We can now express the Stokes operator in any direction ${\bf n}$ as \beq \sind{{\bf n}} = \hat{\mathbf{S}} \cdot \mathbf{n} = \nx
\sx + \ny \sy + \nz \sz. \label{Eq: Linear expansion}\eeq

In addition to the commutation relation (\ref{ccrsu2}), one also has the relation
\begin{equation}
  \label{Eq: S0 commutator}
  [ \sind{0} , \sind{j}] = 0, \quad j \in \{1,2,3\}.
\end{equation}
This indicates that there exist simultaneous eigenstates of $\sind{0}$ (giving the total photon number) and any other Stokes operator. So in principle, a measurement of $\sind{{\bf n}}$, if repeated on many members of an identically prepared ensemble, also allows the photon number statistics to be determined. In fact, a common way to measure any Stokes operator $\sind{{\bf n}} = \hat{U}_{\bf n} \sind{3} \hat{U}_{\bf n}^\dagger$, where $\hat{U}_{\bf n}$ is a (unitary) linear polarization transformation rotating the axis 3 to align with ${\bf n}$, is to first ``rotate'' the state according to $\oprho \rightarrow \hat{U}_{\bf n}^\dagger \oprho \hat{U}_{\bf n}$, and then measure $\sind{3}$. That is, after the rotation of the state, one separates the $H$ and $V$ modes by polarization optics and then counts the number of photons in each mode. The photo-count difference then gives the measured $\sind{{\bf n}}$ eigenvalue, while the sum gives the $\sind{0}$ eigenvalue. This suggests that in a full description of quantum polarization, the excitation manifolds should be treated separately~\cite{Luis2006,Raymer,Bjork}. In consequence, coherences between different manifolds do not carry polarization information. Below we shall use the total photon number $N$ as an index of the excitation manifold. As it will simplify the subsequent discussion, we introduce the normalized $N$-photon density matrix defined as
\beq
 \rho_{mn,N} = \frac{1}{p_N}  \langle m , N - m | \hat{\rho} |n , N - n \rangle, \quad m,n \in \{ 0 , \ldots , N \}, \label{rhoN}
\eeq
where $p_N = \sum_{m=0}^N \langle m , N - m | \hat{\rho} |m , N - m \rangle$.
With this definition, we have
\begin{equation}
\langle \hat{\mathbf{S}}\rangle_N = \Tr ( \oprho_N \hat{\mathbf{S}} ).
\label{Eq: Stokes vector}
\end{equation}
\textit{In reality, it may be experimentally difficult to divide the polarization measures into excitation manifolds, except for few-photon states.} To this end, we shall also define the excitation averaged Stokes vector
\begin{equation}
\langle \hat{\mathbf{S}} \rangle = \sum_{N=1}^\infty p_N \langle \hat{\mathbf{S}}\rangle_N.
\label{Eq: Stokes vector average}
\end{equation}
\textit{All other measures of polarization, defined below, can be averaged over the manifolds in the same manner.} In Secs. \ref{Subsec: coherent} and \ref{Subsec: thermal} this is done when discussing two-mode coherent states and two-mode thermal states, states that when they are used often contains a large, but indeterminate, number of photons. However, for few photon states it is possible to divide the results according to photon number through coincidence measurements, and we believe that it soon will become common to use detectors with photon number resolving capability for such states.

The idea that we will develop below is that the $r$th-order polarization in the $N$th excitation manifold is characterized by a data set that can predict $\langle \sind{{\bf n}}^r \rangle_N$ for any direction of the unit vector ${\bf n}$ on the Poincar\'{e} sphere.

\section{First-order polarization moments}
\label{Sec:First order}

Since the classical description of polarization is based on the first-order moments, the quantum description is the direct translation of the classical description. That is, the Stokes vectors $\langle \hat{\mathbf{S}}\rangle_N$ defined in (\ref{Eq: Stokes vector}) gives the complete first moment polarization information. It follows from the expectation value of both sides of Eq.~(\ref{Eq: Linear expansion}) with regards to the state $\oprho_N$, that $\langle \hat{\mathbf{S}}\rangle_N$ is sufficient to predict $\Tr(\oprho_N \sind{{\bf n}} )$ for any ${\bf n}$.

\section{Assessing the second-order polarization moments}
\label{Sec:Second order}

How should one then go about to characterize higher-order
polarization properties? One way would be to assess all second-order moments, i.e., all polarization correlation values of the form $ T_{j k}^{(2,N)} (\hat{\rho}) = \mathrm{Tr} ( \hat{\rho}_N \hat{S}_{j} \hat{S}_{k} )$, $j,k \in \{1,2,3\}$. However, only when $j=k$ these operator products are
Hermitian, so the expectation values cannot be measured directly. Nonetheless, from a theoretical perspective such an approach is viable and equivalent to the description via polarization moments in different directions. In~\cite{soderholm} we have followed this path. A great simplification and reduction in data is to collect the polarization correlation information into Hermitian moment components~\cite{soderholm}. One of this method's advantage is its simple hierarchy over the moment orders. It is straightforward to understand how to systematically collect the needed, non-redundant information moment order by moment order. Experimentally, it is equivalent to the proposed method in that one measures successively higher moments of the Stokes operator for selected directions on the Poincar\'{e} sphere and then solves an ensuing equation system. A drawback is that it is not so easy to see the relative importance of the moment orders, as lower order moments contribute to higher order moments.

Another method is via the two-mode coherence matrices~\cite{klyshko97} where the $r$th order coherence matrix coefficient $jk$ is defined by $\langle (\hat{a}_H^\dagger)^j (\hat{a}_V^\dagger)^{r-j}\hat{a}_H^k \hat{a}_V^{r-k} \rangle$. This method is informationally equivalent to the method we shall develop. Among the advantages with this method is that all moment coefficients are expectation values of normally ordered annihilation and creation operators, making calculations for coherent states particularly easy. Another advantage is that for $N$-photon states, all coherence matrices of order $r > N$ vanish. However, the method has only an indirect connection to the Stokes operators, and the ensuing matrices give little direct ``feeling'' for the polarization properties of the state, although they contain all the needed data. Experimentally, the off-diagonal coefficients $j \neq k$ of the coherence matrices are not straightforward to measure. In~\cite{Schilling} a method using phase plates and projection onto an $r$th order ``intensity'' of one of the modes, namely $\langle (\hat{a}_H^\dagger)^r \hat{a}_H^r\rangle$, is proposed (note that this is not equal to a measurement of $\langle (\hat{a}_H^\dagger \hat{a}_H)^r\rangle$). Choosing properly $(r+1)^2$ different settings of the phase plates and solving the ensuing set of linear equations the coherence matrix of order $r$ can be obtained. For a state containing up to, and including, $N$ photons, $N(2N^2 +9N + 13)/6$ measurements are thus required using the measurement scheme proposed in~\cite{Schilling}, roughly twice more than for our scheme, see the end of Sec.~\ref{Sec:Fourth and higher order} and Sec.~\ref{Sec: Polarization properties} below. The proposal in~\cite{Schilling} discuss only characterization of $N$-photon states, and not how to assess the polarization or higher order coherence-properties of states with an indeterminate number of photons.

To see how the polarization central moments appear quite naturally in a polarization description, we expand each operator in a state-dependent mean and a fluctuation part, v.i.z. \beq \fluct{\mathbf{n},N}(\oprho) \equiv \sind{\mathbf{n}}
- \Tr ( \oprho_N \sind{\mathbf{n}} ) . \eeq In the following, to simplify the notation, we shall write $\Tr ( \oprho_N \sind{\mathbf{n}}^r ) \equiv \langle \sind{\mathbf{n}}^r \rangle_N$ and $\Tr [ \oprho_N \fluct{\mathbf{n},N}^r(\oprho)] \equiv \langle \fluct{\mathbf{n}}^r \rangle_N$. This allows us to write, for $r=2$
\begin{eqnarray} \langle \sind{{\bf n}}^2 \rangle_N & = & \nx^2
(\langle \sx \rangle_N^2 + \langle \fluct{1}^2\rangle_N)+ {\rm cycl.} + {\rm cycl.}  \nonumber \\
&& + \nx \ny (2\langle \sx\rangle_N \langle \sy\rangle_N + \langle
\fluct{1} \fluct{2} \rangle_N + \langle
\fluct{2} \fluct{1} \rangle_N) \nonumber \\ && + {\rm
cycl.} \nonumber \\ && + {\rm cycl.}, \label{Eq: Expansion}
\end{eqnarray}
where cycl. denote a cyclic permutation of the indices. We see that apart from
$\langle \si \rangle_N$, $j \in \{1,2,3 \}$, the expectation values of the
six Hermitian fluctuation ``operators'' in (\ref{Eq: Expansion}) are
what is needed to know $\langle \sind{{\bf n}}^2 \rangle_N$ in any
direction. These expectation values are the second-order central-moments, (coinciding with the second-order cumulant) defined as
\beq
\langle \fluct{j} \fluct{k} \rangle_N = \langle \sind{j} \sind{k} \rangle_N - \langle \sind{j} \rangle_N \langle  \sind{k} \rangle_N.
\eeq
As can be seen from (\ref{Eq: Expansion}) it is convenient and natural to collect the mixed-product ($j \neq k$) central moments into Hermitian terms, e.g., $\langle  \fluct{j} \fluct{k} + \fluct{k} \fluct{j}\rangle_N$. These terms can be measured, and we see that in addition to the Stokes parameters, we need six more numbers to fully characterize the second-order polarization-properties. The first three can be obtained from measuring the statistics of the Stokes vector $\op{\mathbf{S}}$ yielding the first-order
moments $\langle (\hat{S}_1,\hat{S}_2,\hat{S}_3)\rangle_N $ and the
variances $\langle \fluct{j}^2 \rangle_N$, $j \in \{1,2,3\}$. The additional three numbers can be obtained from
measuring the statistics of $\sind{{\bf n}}$ along the ``diagonal''
directions $(1,1,0)/\sqrt{2}$, $(1,0,1)/\sqrt{2}$, $(0,1,1)/\sqrt{2}$ in the $\sx \sy$, $\sx \sz$, and $\sy \sz$ planes, respectively, corresponding to the
angles $(\theta,\phi)$ of $(\pi/2,\pi/4), (\pi/4,0)$, and
$(\pi/4,\pi/2)$ on the Poincar\'{e} sphere, and then using (\ref{Eq: Expansion}).

As a minor digression, these second-order central-moment terms are directly connected to the Hermitian polarization covariance matrix
$\bm{\Gamma}_N$ with matrix coefficients \beq \Gamma_{j k,N} =
\frac{1}{2} \langle  \fluct{j} \fluct{k} +
\fluct{k} \fluct{j}\rangle_N , \label{Eq:Gamma}\eeq where $j,k \in
\{1,2,3 \}$~\cite{Barakat:1989ys}.
Each such matrix has six independent coefficients as $\Gamma_{j k,N}=\Gamma_{k
j,N}$ by construction. It is clear from Eq.~(\ref{Eq: Expansion}) that this
covariance matrix contains the information we need, in
addition to the expectation value of the Stokes vector, to be able
to predict the value of $\langle \sind{{\bf n}}^2 \rangle_N$ in any
direction. We also have $\langle \fluct{\mathbf{n}}^2 \rangle_N =
\mathbf{n} \cdot \bm{\Gamma}_N\cdot \mathbf{n}^{\rm t}$, where $t$
denotes the transpose.

Every covariance matrix $\bm{\Gamma}_N$ can be made diagonal by an
orthogonal matrix $\mathbf{R}$. In this rotated, orthogonal
coordinate system, where $\sind{\mathbf{e}_j}$ point in the direction of
eigenvector $\mathbf{e}_j$, $j \in \{1,2,3\}$ of $\bm{\Gamma}_N$, one finds the
extreme values of $\langle \fluct{{\bf n}}^2\rangle_N$. In this
coordinate system Eq.~(\ref{Eq: Expansion}) simplifies to \beq
\langle \fluct{\mathbf{n}}^2 \rangle_N =  \eigen_1 (\sin \theta' \cos
\phi')^2  + \eigen_2 (\sin \theta' \sin \phi')^2 + \eigen_3
\cos^2 \theta' , \eeq where $\eigen_j$ is the $j$th eigenvalue of
$\bm{\Gamma}_N$, $\theta'$ is the angle between $\mathbf{n}$ and $\mathbf{e}_3$, and $\phi'$ is the azimuthal angle in the $\mathbf{e}_1$-$\mathbf{e}_2$ plane. This equation may look like the equation of an ellipsoid, but it is not, as this is the magnitude of the variance
of $\langle \fluct{\mathbf{n}}^2 \rangle$ {\it in the direction}
$\mathbf{n}$ on the Poincar\'{e} sphere.

In order to measure $\bm{\Gamma}_N$, one makes the same measurements as were discussed above. The matrix
$\bm{\Gamma}_N$ can subsequently be deduced by solving Eq.~(\ref{Eq: Expansion}) for
$\langle \sind{j} \rangle_N$ and $\langle \fluct{j}^2 \rangle_N$ given the measured values of $\langle \sind{\mathbf{n}}^2 \rangle_N$ along the six directions. For better ``immunity'' to systematic errors,
one could make measurements along additional directions and
subsequently make a best fit of the ensuing overcomplete system of
equations.

\section{Third-order polarization}
\label{Sec:Third order}
Moving on to third-order moments, things get a bit more involved. Still,
our underlying idea is that if one has all the central moments up to order three, then one can predict $\langle \sind{{\bf
n}}^3 \rangle_N$ for any direction.

We therefore first express the expectation value $\langle \hat{S}_{\mathbf{n}}
\rangle_N^3$ in terms of $\langle {\bf \hat{S}} \rangle_N$:
\begin{eqnarray}
\langle \sind{\mathbf{n}} \rangle_N^3 & = & \nx^3 \langle \sx \rangle_N^3 + \ny^3 \langle \sy \rangle_N^3 + \nz^3 \langle \sz \rangle_N^3 \nonumber \\
& & + 3 \mbox{{\Large (}} \nx^2 \ny \langle \sx \rangle_N^2 \langle \sy \rangle_N + \nx^2 \nz \langle \sx \rangle_N^2 \langle \sz \rangle_N \nonumber \\
& & + \ny^2 \nx \langle \sy \rangle_N^2 \langle \sx \rangle_N + \ny^2 \nz \langle \sy \rangle_N^2 \langle \sz \rangle_N \nonumber \\
& & + \nz^2 \nx \langle \sz \rangle_N^2 \langle \sx \rangle_N + \nz^2 \ny \langle \sz \rangle_N^2 \langle \sy \rangle_N \mbox{{\Large )}} \nonumber \\
& & + 6 \nx \ny \nz \langle \sx \rangle_N \langle \sy
\rangle_N \langle \sz \rangle_N.
\label{Eq: Mean expansion third order}
\end{eqnarray}
In a similar manner we can  express the third-order raw moment of $\sind{\mathbf{n}}$ as
\begin{eqnarray}
\langle \sind{\mathbf{n}}^3 \rangle_N & = & \nx^3\left ( \langle \fluct{1}^3 \rangle_N + 3 \langle \sx \rangle_N \langle \fluct{1}^2 \rangle_N  \right ) \nonumber \\ & & + {\rm cycl.} + {\rm cycl.} \nonumber \\
& & + \nx^2 \ny\mbox{{\Large (}}  3 \langle \sx \rangle_N \langle \fluct{1} \fluct{2} + \fluct{2} \fluct{1}\rangle_N \nonumber \\
& & + 3 \langle \sy \rangle_N \langle \fluct{1}^2 \rangle_N +
\langle \fluct{1}^2 \fluct{2} + \fluct{2} \fluct{1}^2\rangle_N \nonumber \\
& & + \langle \fluct{1} \fluct{2} \fluct{1} \rangle_N \mbox{{\Large )}}+ {\rm cycl.} + {\rm cycl.} \nonumber \\
& & + \nx^2 \nz\mbox{{\Large (}}  3 \langle \sx \rangle_N \langle \fluct{1} \fluct{3} + \fluct{3} \fluct{1}\rangle_N \nonumber \\
& & + 3 \langle \sz \rangle_N \langle \fluct{1}^2 \rangle_N +
\langle \fluct{1}^2 \fluct{3} + \fluct{3} \fluct{1}^2\rangle_N \nonumber \\
& & + \langle \fluct{1} \fluct{3} \fluct{1} \rangle_N \mbox{{\Large )}}+ {\rm cykl.} + {\rm cykl.} \nonumber \\
& & +  \nx \ny \nz \mbox{{\Large (}} 3\langle \sx \rangle_N
\langle \fluct{2}
\fluct{3} + \fluct{3} \fluct{2}\rangle_N \nonumber \\
& & + \langle \fluct{1}\fluct{2} \fluct{3} + \fluct{1}\fluct{3} \fluct{2}\rangle_N \nonumber \\
& & + {\rm cycl.} + {\rm cycl.} + \langle \sind{\mathbf{n}} \rangle_N^3 \mbox{{\Large )}}. \label{Eq: Third
order expansion}
\end{eqnarray}
Finally we can express the third-order central-moments as
\begin{eqnarray}
\langle \fluct{j} \fluct{k} \fluct{\ell} \rangle_N & = & \langle \sind{j} \sind{k} \sind{\ell}\rangle_N - \langle \sind{j} \rangle_N \langle  \sind{k} \sind{\ell} \rangle_N \nonumber \\
&& - \langle \sind{k} \rangle_N \langle  \sind{j} \sind{\ell} \rangle_N  - \langle \sind{\ell} \rangle_N \langle  \sind{j} \sind{k} \rangle_N \nonumber \\
&& + 2 \langle \sind{j} \rangle_N \langle  \sind{k} \rangle_N \langle \sind{\ell} \rangle_N .
\end{eqnarray} Hence, for the first to third order, the central moments coincide with the cumulants.

One sees that in (\ref{Eq: Third
order expansion}), if the ten Hermitian, third order, central moment terms, each associated
with a different geometric term $n_j n_k n_{3-j-k}$, $j,k \in \{1,2,3\}$, $j+k \leq 3$ are
determined, in addition to the first and second-order properties, then
the third-order polarization-properties are also determined for any direction. Hence, what one needs to measure are the sums of all fluctuation terms having $j$ ones, $k$ twos, and $3-j-k$ threes, where $j + k \leq 3$, or more generally, for polarization order $r$, into sums having $r-j-k$ threes, where $j + k \leq r$.

Measuring the third-order fluctuations along, e.g., the $(\theta,\phi)$
directions  $(0,0)$, $(\pi/2,0)$, $(\pi/2,\pi/2)$,
$(\pi/2,\phi_1)$, $(\pi/2,-\phi_1)$, $(\pi/2-\phi_1,0)$,
$(\pi/2+\phi_1,0)$, $(\pi/2-\phi_1,\pi/2)$,
$(\pi/2+\phi_1,\pi/2)$, and $(\pi/2-\phi_1,\pi/4)$, where
$\phi_1 = \arccos \sqrt{2/3}$, one gets a system of ten linearly
independent equations that allows one to determine the terms
$\langle \fluct{1}^3 \rangle_N + 3 \langle \sx \rangle_N \langle
\fluct{1}^2 \rangle_N$ etc. Using the knowledge about the lower-order polarization terms, one can subsequently estimate the third-order terms, in this case $\langle \fluct{1}^3 \rangle_N$. We note that the three first measurement directions are simply along the $\sx$, $\sy$ and $\sz$ axes, so in fact, only measurement along seven extra directions are needed, in addition to the measurements along six directions needed to determine $\langle \mathbf{\hat{S}} \rangle_N$ and $\langle \sind{\mathbf{n}}^2\rangle_N$. Alternatively, if one wants to minimize the number of measurement directions, one can use the statistics collected when measuring along the six directions that determine the first and second-order polarization moments, and supplement them with measurements along the four new directions $(\pi/6,\pi/6)$, $(\pi/6,\pi/3)$, $(\pi/3,\pi/6)$, $(\pi/3,\pi/3)$.

For third-order polarization the first thing to be considered is
that the fluctuations of $\hat{S}_\mathbf{n}^3 $ involves not only third
powers of $\fluct{j}$, but also terms like $\nx^3 \langle \sx
\rangle \langle \fluct{1}^2 \rangle_N$ and $\nx^2 \ny \langle \sx
\rangle_N \langle \fluct{1} \fluct{2} + \fluct{2} \fluct{1} \rangle_N$.
That is, the second and the third-order fluctuations become
``intermixed'' in this polarization order unless the state has
vanishing Stokes parameters. This is in contrast to the (simpler) second
order. A consequence of this is that if the state's first-order
polarization is much larger than the square root of its variance, then all higher-order fluctuations will, in general, be dominated
by the beating terms between the mean polarization vector and the
second-order fluctuations. Hence, for
most ``reasonably excited'' and ``somewhat first-order polarized''
states {\it one needs not go beyond the second-order moments to
characterize the polarization fluctuations of all orders to a very good
precision.} However, for states having a small or vanishing first
order polarization, and for, e.g., the eigenstates to the Stokes operators
in the direction of $\bm{\Gamma}_N$'s eigenvector
directions on the Poincar\'{e} sphere, the polarization structures
of orders higher than two will be of interest.

The expansions (\ref{Eq: Expansion}) and (\ref{Eq: Third order expansion}) also indicate an experimental advantage in describing the polarization in terms of increasing orders of its central moments. For each order it becomes quite clear to which accuracy one needs to measure the moments to obtain information not already contained in lower moments and similarly, to what extent the higher-order central moments contribute to the raw moments. This information is of course implicit in ``equivalent'' descriptions such as generalized coherence matrices~\cite{klyshko97} or polarization tensors~\cite{soderholm}, but it is not explicitly displayed.

\section{Fourth- and higher-order polarization}
\label{Sec:Fourth and higher order}

From the preceding sections it is rather clear how one could continue through the higher orders. In order to know $\langle \sind{\mathbf{n}}^r \rangle_N$ in any direction, the full set of Hermitian central-moment terms for all orders $\leq r$ is needed.

Explicitly, the fourth order the central moment is

\begin{widetext}
\begin{eqnarray}
\langle \fluct{j} \fluct{k} \fluct{\ell} \fluct{m} \rangle_N & = & \langle \sind{j} \sind{k} \sind{\ell} \sind{m}\rangle_N - \langle \sind{j} \rangle_N \langle  \sind{k} \sind{\ell} \sind{m} \rangle_N - \langle \sind{k} \rangle_N \langle  \sind{j} \sind{\ell} \sind{m}\rangle_N  - \langle \sind{\ell} \rangle_N \langle  \sind{j} \sind{k} \sind{m}\rangle_N \nonumber \\
&&-\langle \sind{m} \rangle_N \langle  \sind{j} \sind{k} \sind{\ell}  \rangle_N + \langle \sind{j} \rangle_N \langle \sind{k} \rangle_N \langle \sind{\ell} \sind{m}\rangle_N + \langle \sind{j} \rangle_N \langle \sind{\ell} \rangle_N \langle \sind{k} \sind{m}\rangle_N + \langle \sind{j} \rangle_N \langle \sind{m} \rangle_N \langle \sind{k} \sind{\ell}\rangle_N\nonumber \\
&& + \langle \sind{k} \rangle_N \langle \sind{\ell} \rangle_N \langle \sind{j} \sind{m}\rangle_N + \langle \sind{k} \rangle_N \langle \sind{m} \rangle_N \langle \sind{j} \sind{\ell}\rangle_N + \langle \sind{\ell} \rangle_N \langle \sind{m} \rangle_N \langle \sind{j} \sind{k}\rangle_N \nonumber \\
&& - 3 \langle \sind{j} \rangle_N \langle  \sind{k} \rangle_N \langle \sind{\ell} \rangle_N \langle \sind{m} \rangle_N . \label{Eq: fourth order}
\end{eqnarray} \end{widetext} In contrast to the three lower orders, this result is not identical to the fourth-order cumulant. Higher-order central moments, that we will not write out explicitly, do not coincide with the cumulants either.

In analogy with the second and third order, we need not determine each term of the form (\ref{Eq: fourth order}), but only the Hermitian sum of moments associated to a certain geometrical pre-factor. The number of such central-moment sum-terms specific to the order $r$ is $(r+1)(r+2)/2$ and the complete set of such terms up to, and including, order $r$ is $r(r^2 + 6r + 11)/6$. To obtain the terms, one would have to measure the polarization statistics along such a number of carefully selected directions, yielding a complete,``maximally'' linearly independent set of equations that could be solved numerically. To obtain better accuracy one could ``oversample'' the polarization statistics over the Poincar\'{e} sphere and use maximum likelihood or entropy methods to make a better estimate. However, as the states that have their polarization characteristics mainly determined by the $r$th-order moment will be rather elaborated as $r$ increases, the interest in the polarization central moment terms will be limited to $r \leq 4$, or so.

\section{Polarization properties and excitation manifolds}
\label{Sec: Polarization properties}
Using the bosonic commutation relation (\ref{BosonicCommutator}), it is possible to rewrite any $r$th-order product of Stokes operators to a sum of normally ordered creation and annihilation operators of maximum order $r$ in the annihilation orders~\cite{Schilling}. To exemplify, one can write \begin{eqnarray} \sz^2 & = & \ahd \ahd \ah \ah - 2 \ahd \avd \ah \av + \avd \avd \av \av \nonumber \\
& & + \ahd \ah + \avd \av . \end{eqnarray}  As all Stokes operators are composed of terms with one creation and one annihilation operator, this implies that all polarization properties of a state with no excitation above the $N$-photon manifold are determined by the polarization moments up the $r=N$th order. All moments of order higher than $N$ can have only those normal ordered terms less or equal to the $N$th order different from zero, and those terms will always be contained in the moments up to, and including, the $N$th order. Below we shall see a specific example of this, namely that for a three-photon state, it is sufficient to require that $\langle \sind{\mathbf{n}}^m \rangle$ is isotropic for $m=1,2,3$ in order for the state to be unpolarized to all orders. Note, however, that should the higher-order central moments be zero, this does not indicate that the state lacks higher-order polarization-structure. Instead, the implication is that this structure can be derived from the ``beating'' terms from lower-order polarizationmoments, as already hinted in Sec.~\ref{Sec:Third order}.

Another consequence of the fact that states with no excitation above the $N$-photon manifold has its polarization fully characterized by its $N$ lowest-order moments is that polarization tomography of such a state is requiring considerably less resources than a full state tomography. For a full state tomography involving the  $(N+1)(N+2)/2$ basis states (e.g., for $N=1$ the states $\ket{0,0}$, $\ket{0,1}$, and $\ket{1,0}$ can be chosen) the density matrix is characterized by $N(N^3 + 6 N^2 + 13 N + 12)/4$ independent real numbers. This can be compared to the $N(N^2 + 6N + 11)/6$ numbers needed for the polarization tomography of such a state. Raymer \textit{et al.} has used the term ``polarization sector'' of the density matrix for the subset of information needed to characterize only a state's polarization~\cite{Raymer}.

This said, an $N$-photon state is fully described by $N(N+2)$ real numbers while the polarization central moments up to, and including the $r=N$:th order require $N(N^2 + 6N + 11)/6$ numbers. Very recently we found a method to reduce the number of measurements to the minimum $N(N+2)$, but the analysis and description of this scheme will be published elsewhere~\cite{soderholm}.

\section{Application to different polarization states}
\label{Sec: Menagerie}

We shall now apply the characterization developed above to a few examples and also compare the theory with experiments in the two-photon excitation manifold. We remind the reader that we use the $\sz$ eigenstates as our basis states. The experimental setup is discussed, and measurement data are given, in the Appendix.

\subsection{SU(2) coherent states}
\begin{figure}
 \includegraphics[width=0.95\columnwidth]{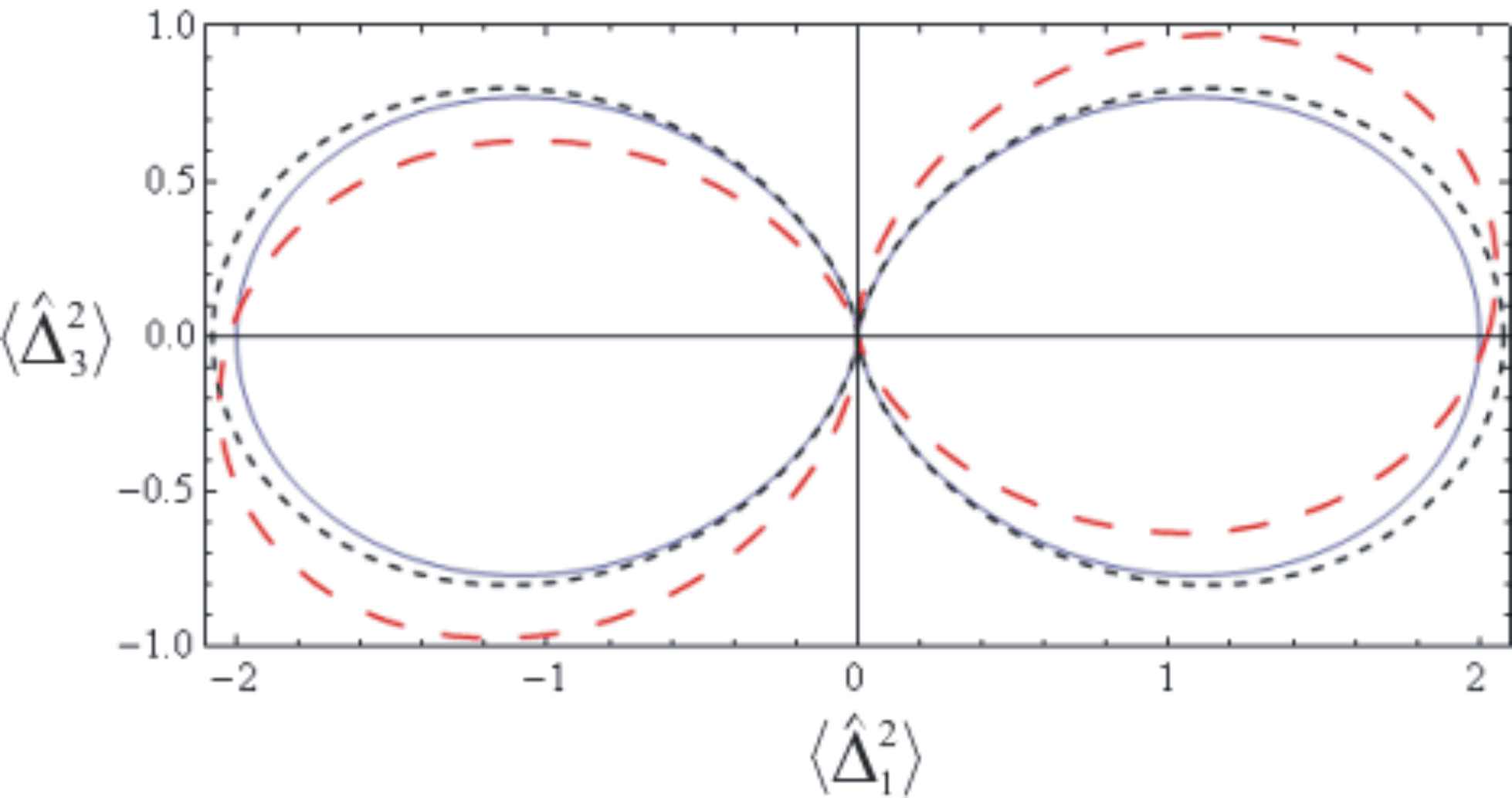}
 \caption{$\langle \hat{\Delta}_\mathbf{n}^2 \rangle$ in the plane defined by $\langle \fluct{2}^2\rangle=0$ for the state $\ket{2,0}$. The theoretically expected result is drawn solid (blue), the experimentally obtained result is drawn dashed (red), and the experimental result, solidly rotated so that its eigenvectors coincide with the intended eigenvectors, is drawn dotted (black).}
 \label{2Hcut}
\end{figure}

Through an appropriate polarization transformation of the state $\ket{N}_H\otimes \ket{0}_V \equiv \ket{N,0}$  any $N$-photon, SU(2) coherent state can be obtained. Since a polarization transformation is equivalent to a rotation of the Poincar\'{e} sphere, it thus suffices to study the state $\ket{N,0}$. Quite clearly, all its moments are zero except in excitation manifold $N$, and therefore we will suppress this index. The state has $\langle \szero \rangle = N$, the Stokes vector is $(0,0,N)$, $\langle \fluct{3}^m \rangle =0$ $\forall \  m$, $\langle \fluct{1}^m \rangle = \langle \fluct{2}^m \rangle =0$ for odd $m$, $\langle \fluct{1}^2 \rangle = \langle \fluct{2}^2 \rangle = N$, and $\langle \fluct{1}\fluct{2} + \fluct{2}\fluct{1}\rangle = \langle \fluct{1}\fluct{3} + \fluct{3}\fluct{1} \rangle = \langle \fluct{2}\fluct{3} + \fluct{3}\fluct{2} \rangle = 0$. Its second-order, polarization central moments are hence reduced to a toroidal structure with radius $N$, with its ``hole'' in the $\sz$ direction on the Poincar\'{e} sphere. Its third-order  central moments have the non-vanishing terms $\langle \fluct{j}^2 \fluct{3} + \fluct{j} \fluct{3} \fluct{j} +\fluct{3}\fluct{j}^2 \rangle= - 2 N$, $j \in \{1,2\}$. The non-vanishing terms of fourth order are $\langle \fluct{j}^4 \rangle=3N^2 - 2N$, $\langle \fluct{j}^2 \fluct{3}^2 + \fluct{3}^2\fluct{j}^2  + \fluct{j}\fluct{3}\fluct{j} \fluct{3} + \fluct{3}\fluct{j}\fluct{3}\fluct{j} + \fluct{j}\fluct{3}^2\fluct{j} + \fluct{3}\fluct{j}^2\fluct{3}\rangle= 4 N$, and $\langle \fluct{1}^2 \fluct{2}^2 + \fluct{2}^2\fluct{1}^2  + \fluct{1}\fluct{2}\fluct{1} \fluct{2} + \fluct{2}\fluct{1}\fluct{2}\fluct{1} + \fluct{1}\fluct{2}^2\fluct{1} + \fluct{2}\fluct{1}^2\fluct{2}\rangle= 6 N^2 - 4 N$, $j \in \{1,2\}$.
\begin{figure}
 \includegraphics[width=0.95\columnwidth]{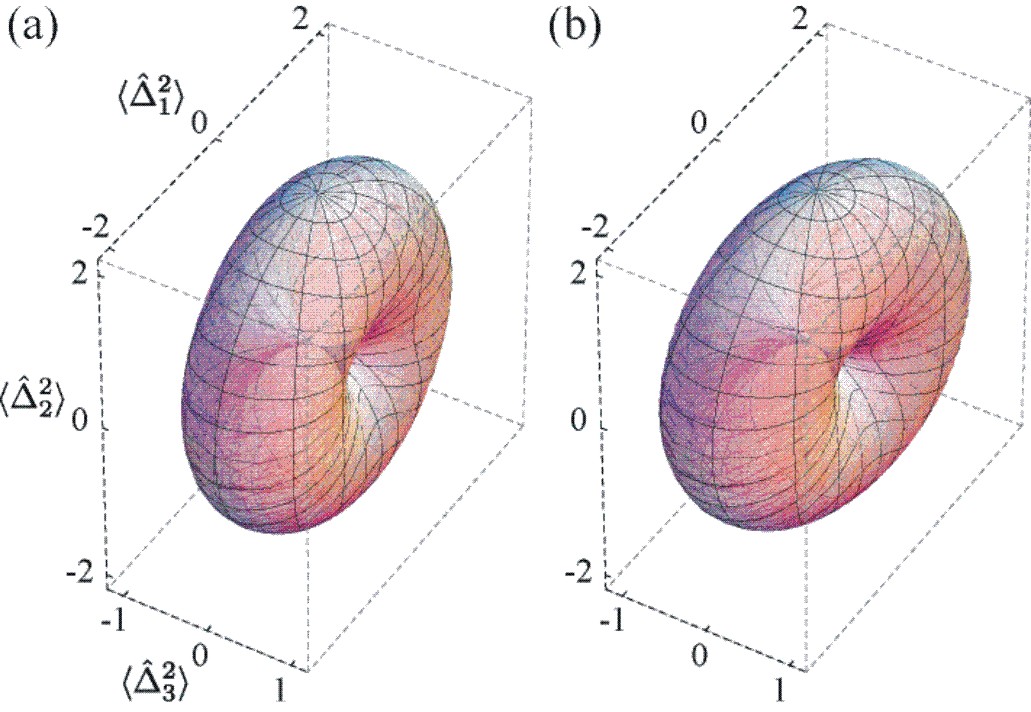}
 \caption{The second order central moment $\langle \hat{\Delta}_\mathbf{n}^2 \rangle$ for the state $\ket{2,0}$. Theoretical plot in (a) and the experimental results in (b).}
 \label{2H}
\end{figure}
This is a minimum-sum uncertainty-state [saturating the left inequality in Eq.~(\ref{Eq: Uncert relation})]. In Fig.~\ref{2H} we plot the theoretically computed function $\langle \fluct{\mathbf{n}}^2 \rangle_2$ for the state $\ket{2,0}$ to the left, and the experimentally obtained results on the right. The measured Stokes vector of this state is $(-0.19, 0.12, 1.97)$. The main source of error in the estimation of the Stokes vector is neither fluctuations nor random errors, but the fact that the generated state and the measurement axes are slightly rotated relative to each other. This is better seen in Fig.~\ref{2Hcut} where we have plotted the theoretically expected result (solid) and the experimental figure (dashed) as derived directly from the measured parameters in Table \ref{Table20} (no fitting done) in the $\sx$-$\sz$ plane. We have subsequently used the measurement data and computed the eigenvectors and eigenvalues for the matrix $\bm{\Gamma}_2$ defined in Eq.~(\ref{Eq:Gamma}). The eigenvalues (that due to the Hermiticity of $\bm{\Gamma}_2$ are the plot's extrema which are found at orthogonal points on the Poincar\'{e} sphere) are 2.08, 2.02, and 0.00, respectively, and, e.g., the eigenvector corresponding to the smallest eigenvalue is $(-0.15, -0.02, 0.99)$. This eigenvector is rotated about 8.1 degrees from the $\sz$ axis, mainly around the $\sy$ axis (roughly in agreement with the orientation of the state's Stokes vector). If we re-plot the figure, with no other ``fitting" than a solid rotation of the experimental figure to make its eigenvectors coincide with the Poincar\'{e} sphere coordinate axes, we obtain the dotted curve in Fig.~ \ref{2Hcut}. Note that the needed rotation is not exactly perpendicular to the drawn plane, so the rotated figure will also change shape slightly. The dotted figure, (and the values of the $\bm{\Gamma}_2$ eigenvalues) confirm that the experimental errors due to fluctuations are below $\pm$4\%. The main errors are systematic, due to imperfect polarization optics and beam splitters. The systematic errors will persist even if additional points on the Poincar\'{e} sphere are measured, indicating that the proposed method is ``efficient'' from a data collecting point of view. By measuring the nine data in Table \ref{Table20} one can hence estimate both the accuracy and the precision of the experiment.

\subsection{$\ket{N,N}$ states}
This state has $\langle \szero \rangle = 2 N$, the Stokes vector is $(0,0,0)$, and $\langle \fluct{3}^m \rangle =0$ $\forall \  m$. The only non-vanishing second-order, central-moment terms are $\langle \fluct{j}^2 \rangle = 2 N (N+1)$, $j \in \ \{1,2\}$. The state has vanishing third-order central-moment in every direction, and its fourth-order, non-vanishing central-moment terms are $\langle \fluct{j}^4 \rangle=2 N (3 N^3 + 6 N^2 + N -2)$, $\langle \fluct{j}^2 \fluct{3}^2 + \fluct{3}^2\fluct{j}^2  + \fluct{j}\fluct{3}\fluct{j} \fluct{3} + \fluct{3}\fluct{j}\fluct{3}\fluct{j} + \fluct{j}\fluct{3}^2\fluct{j} + \fluct{3}\fluct{j}^2\fluct{3}\rangle= 8 N (N+1)$, and $\langle \fluct{1}^2 \fluct{2}^2 + \fluct{2}^2\fluct{1}^2  + \fluct{1}\fluct{2}\fluct{1} \fluct{2} + \fluct{2}\fluct{1}\fluct{2}\fluct{1} + \fluct{1}\fluct{2}^2\fluct{1} + \fluct{2}\fluct{1}^2\fluct{2}\rangle= 4 N (3N^3 + 6N^2 + N - 2)$, $j \in \{1,2\}$. This is a pure, maximum-uncertainty state, saturating the right inequality in Eq.~(\ref{Eq: Uncert relation})]. If one plots the experimentally obtained results one obtains a figure very similar to Fig.~\ref{2H} (b), except that $\langle \fluct{\mathbf{n}}^2 \rangle_2$ for this state is approximately twice as large in all directions as for the state $\ket{2,0}$.   The measured Stokes vector of this state is $(-0.01, -0.08, 0.01)$. Extracting the measured state's extremal values for $\langle \fluct{\mathbf{n}}^2 \rangle_2$ from the experimental data (i.e. the eigenvalues of $\bm{\Gamma}_2$) one obtains 4.10, 3.98, and -0.03, indicating an experimental error due to fluctuations below $\pm$3\%. This data too, indicates a slight mismatch (about 12 degrees) between the state's intended orientation on the Poincar\'{e} sphere and its measured orientation. Note that the state's orientation cannot be obtained from the Stokes vector, since nominally it vanishes. The the measured values are simply the measurement errors. However, from the second-order moment data the state's orientation relative to the Poincar\'{e} sphere axes can be well determined.

\subsection{Two-mode coherent states}
\label{Subsec: coherent}
Any two-mode, coherent state $\ket{\alpha',\alpha''}$ can be converted into the state $\ket{|\alpha|,0}$, where $|\alpha|^2 = |\alpha'|^2 + |\alpha''|^2$, by a polarization transformation. Therefore it suffices to study the latter state, which can be written \beq \exp(-|\alpha|^2/2) \sum_{N=0}^\infty \frac{|\alpha|^N}{\sqrt{N!}} \ket{N,0}. \eeq In each excitation manifold except the non-excited one, the state has the same central moments as a SU(2) coherent state. Summing over the manifolds, the coherent state has $\langle \szero \rangle = |\alpha|^2$, the Stokes vector $(0,0,|\alpha|^2)$ and $\langle \fluct{j}^2 \rangle =|\alpha|^2$ for $j \in \{1,2,3\}$. The off-diagonal coefficients of the covariance matrix $\bm{\Gamma}$ are zero, so the second-order central moment is isotropic with radius $|\alpha|^2$. In the third order, the only non-vanishing central-moment terms are: $\langle \fluct{3}^3 \rangle = \langle \fluct{1}^2 \fluct{3} + \fluct{1}\fluct{3}\fluct{1} + \fluct{3}\fluct{1}^2\rangle = \langle \fluct{2}^2 \fluct{3} + \fluct{2}\fluct{3}\fluct{2} + \fluct{3}\fluct{2}^2\rangle = |\alpha|^2$. The non-vanishing fourth-order central-moment terms are:
\beq
\langle \fluct{j}^4 \rangle =3|\alpha|^4 + |\alpha|^2 , \nonumber \eeq for $j \in \{1,2,3\}$, and
\begin{eqnarray} \langle \fluct{j}^2\fluct{k}^2 + \fluct{j}\fluct{k}\fluct{j}\fluct{k} + \fluct{j}\fluct{k}^2\fluct{j} + \fluct{k}\fluct{j}^2\fluct{k} \nonumber \\ + \fluct{k}\fluct{j}\fluct{k}\fluct{j} +  \fluct{k}^2\fluct{j}^2\rangle = 6 |\alpha|^4 + 2 |\alpha|^2 , \nonumber \end{eqnarray}
for $j,k \in \{1,2,3\}$ and $j<k$.

\subsection{Unpolarized states, SU(2) invariant states, and thermal states}
\label{Subsec: thermal}
The unpolarized states~\cite{soderholm,Bjork} have isotropic central moments for all orders. Due to symmetry, the odd order central moments are identically zero. For an $N$-photon unpolarized state, the second-order central-moment in any direction is $\langle \fluct{{\bf n}}^2 \rangle = \langle \sind{{\bf n}}^2 \rangle = N(N+2)/3$, and the fourth-order central moment is $\langle \fluct{{\bf n}}^4 \rangle = N(N+2)(3 N^2 + 6 N -4)/15$.

SU(2) invariant states~\cite{prakash} constitute a subclass of the unpolarized states~\cite{Bjork}. They take the form
\beq
 \sum_{N=0}^\infty \frac{p_N \hat{\openone}_N}{N+1} ,
\eeq
where $p_N$ is a probability distribution and $\hat{\openone}_N$ is the identity operator in the $N$th manifold.
A simple example of a state that is unpolarized but not SU(2) invariant is given in Ref.~\cite{soderholm}. In Table \ref{Table:Unpolarized} we have tabulated the measured data for the  unpolarized (and SU(2) invariant) two-photon state $(\ket{2,0}\bra{2,0} + \ket{1,1}\bra{1,1} + \ket{0,2}\bra{0,2})/3$. The measured Stokes vector for this state is $(-0.07,-0.10,0.01)$. As can be read essentially directly from the table, the experimentally obtained function $\langle \fluct{{\bf n}}^2 \rangle$ is more or less a sphere. The experimentally obtained extremal values of the function are 2.72, 2.68, and 2.62, indicating an error due to fluctuations of about $\pm$ 2\%.

The thermal states, finally, constitute a subclass of the SU(2) invariant states with $p_N=\bar{N}^N/(1+\bar{N})^{N+1}$, where $\bar{N}$ is the average excitation $[\exp(h \nu/k T)-1]^{-1}$, where $h$ is Planck's constant, $\nu$ is the optical frequency, $k$ Boltzmann's constant, and $T$ the temperature. Their second- and fourth-order central moments are given above, and their excitation probability averaged second- and fourth-order central moments are $\bar{N} + 2 \bar{N}^2/3$ and $\bar{N} + 26 \bar{N}^2/3 + 12 \bar{N}^3+24 \bar{N}^4/5$, respectively.

\subsection{A two-photon mixed state}
\begin{figure}
 \includegraphics[width=0.5\columnwidth]{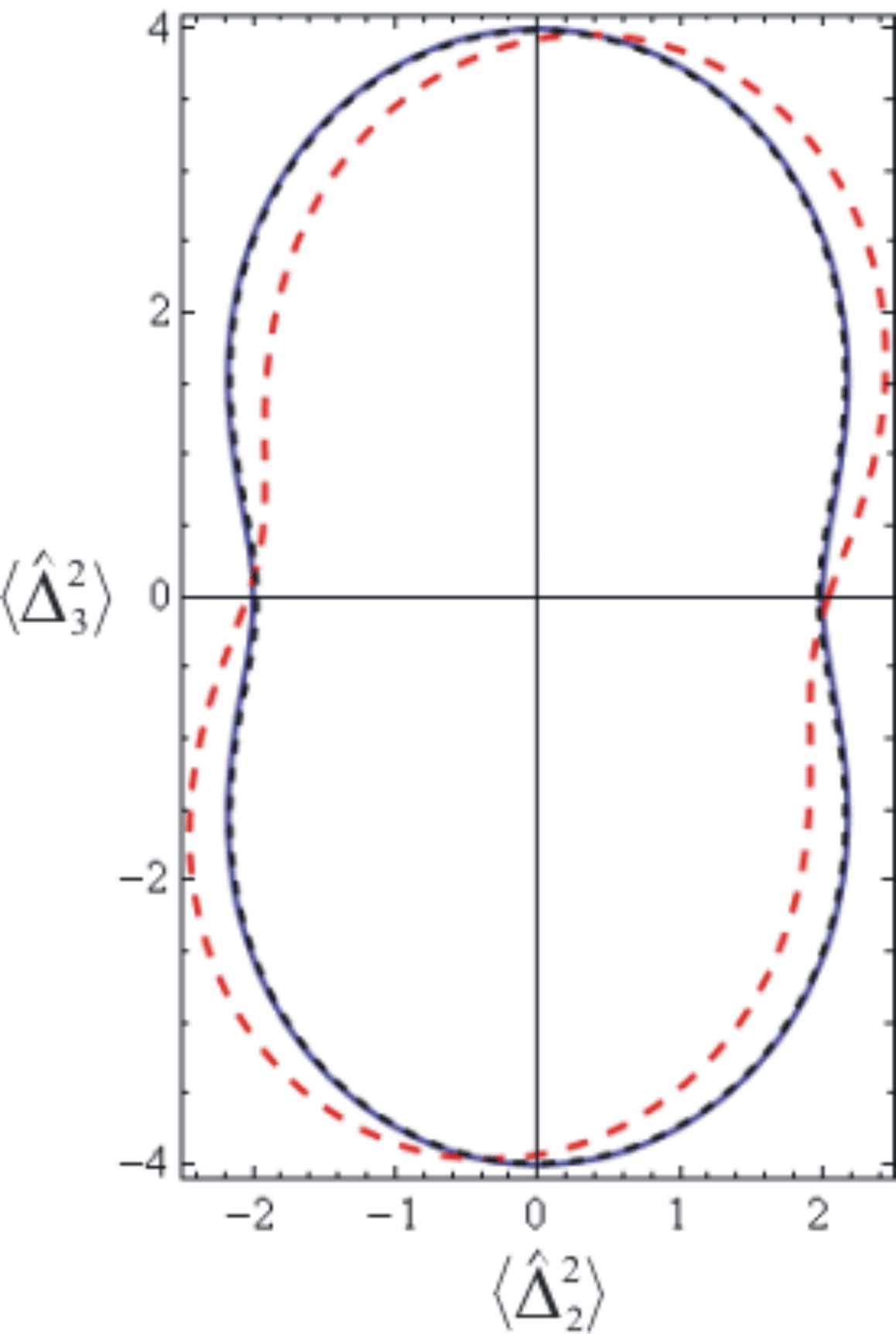}
 \caption{$\langle \hat{\Delta}_\mathbf{n}^2 \rangle$ in the plane defined by $\langle \fluct{1}^2\rangle=0$ for the state $(\ket{2,0}\bra{2,0} +  \ket{0,2}\bra{0,2})/2$. The theoretically expected result is drawn solid (blue), the experimentally obtained result is drawn dashed (red), and the experimental result, solidly rotated so that its eigenvectors coincide with the intended eigenvectors, is drawn dotted (black).}
 \label{Cutmixture}
\end{figure}

By mixing states, one can get quite complicated polarization characteristics. Below, we shall elaborate on this for three photon states. For two-photon states the parameter space is of course smaller. For example, for the state $(\ket{2,0}\bra{2,0} +  \ket{0,2}\bra{0,2})/2$, the second-order central-moment $\langle \fluct{\mathbf{n}}^2 \rangle_2$  depicted in Fig.~\ref{2H,2V_mixture} (theory on the left, experiments on the right). The measured Stokes vector of this state is $(-0.11,-0.10,0.00)$.  \begin{figure}
 \includegraphics[width=0.6\columnwidth]{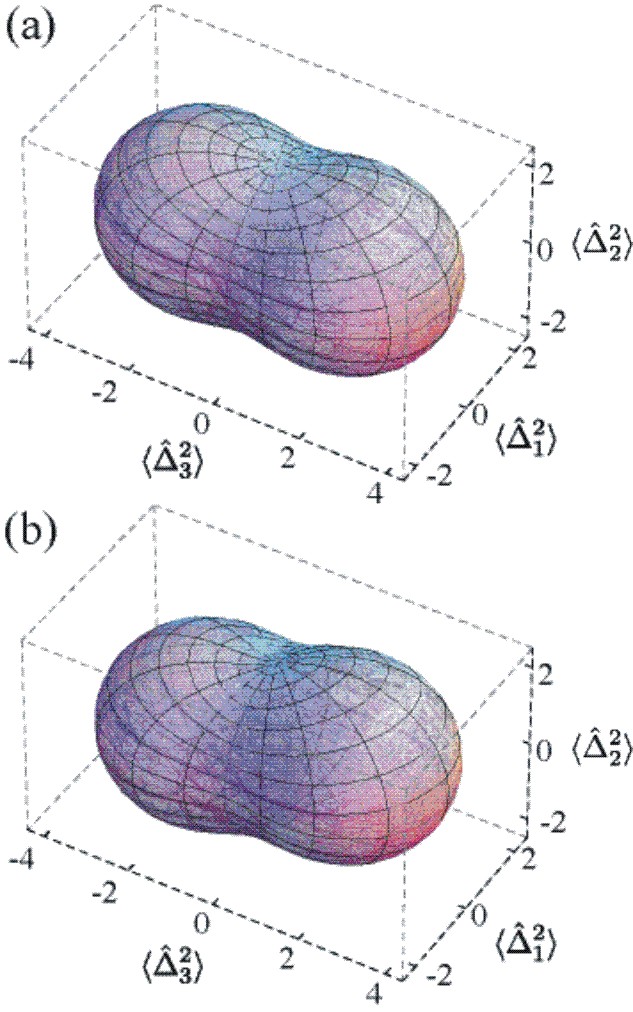}
 \caption{The second-order polarization central moment $\langle \hat{\Delta}_\mathbf{n}^2 \rangle$ for the state $(\ket{2,0}\bra{2,0} +  \ket{0,2}\bra{0,2})/2$. Theoretical plot in (a) and the experimental results in (b).}
 \label{2H,2V_mixture}
\end{figure}
To better appreciate the random and the systematic errors of the measurement, a cut through Fig.~\ref{2H,2V_mixture} in the $\langle \fluct{1}^2\rangle=0$ plane is shown in Fig.~\ref{Cutmixture}. Again it is seen that the experimentally obtained figure is not aligned with the intended orientation, but that the experimentally obtained figure is rotated about 10 degrees around the $\sx$ axis. The figure's extremal values are 3.99, 2.03, and 1.97. Reorienting the figure by a solid rotation so that the measured and intended axes coincide, one obtains the dotted curve in Fig.~\ref{Cutmixture}. By so removing the systematic errors, one obtains a very good ($\pm$2\%) agreement between the experiments and the theory.

\subsection{Some 3-photon states}

Here we explore the polarization characteristics of a few
states up to the third order and show
that for $N=3$ there exist six classes of states if they are sorted according
to their first, second, and third-order polarization central moments. Before giving examples of the classes, it is helpful to retain the
uncertainty relation (\ref{Eq: Uncert relation}). In the third excitation
manifold this means that the sum of the second-order polarization
variances must lie between the values 6 and 15. In order to have an isotropic
second-order central moment, the diagonalized covariance matrix $\bm{\Gamma}$
should be proportional to the $3 \times 3$ unit matrix, and the
relation above dictates that the proportionality factor must be in
the range between 2 to 5. In fact, only minimum sum uncertainty states will
reach the lower limit in (\ref{Eq: Uncert relation}) and such
states have an anisotropic second-order polarization central moment. We conjecture that the lower limit for an isotropic second-order central moment is in fact $\langle \szero \rangle$ so that the minimum uncertainty sum for second order isotropic states is $3 \langle \szero \rangle$ (and specifically 9 for three-photon states). The corresponding state is $\ket{N,0}\bra{N,0}(1 \pm [\{N-1\}/N]^{1/2})/2 + \ket{0,N}\bra{0,N}(1 \mp [\{N-1\}/N]^{1/2})/2$, and in this specific manifold $(1/2 \pm 6^{-1/2})\ket{3,0}\bra{3,0}+(1/2 \mp 6^{-1/2})\ket{0,3}\bra{0,3}$.

Since states that lack first-order polarization but are second order polarized have already been discussed, we shall now look at states that have an isotropic, second-order polarization central moment, but that may have higher-order polarization-structure. Applying the requirements for a state to have isotropic polarization up to second order, one can derive such a three-photon state's density matrix $\oprho$ to be of the form
\beq \oprho = \left(
               \begin{array}{cccc}
                 \varrho_{11} & \varrho_{12} & \varrho_{13} & \varrho_{14} \\
                 \varrho_{12}^* & 1-3\varrho_{11} & -\sqrt{3} \varrho_{12} & -\varrho_{13} \\
                 \varrho_{13}^* & -\sqrt{3} \varrho_{12}^* & 3 \varrho_{11}-\frac{1}{2} & \varrho_{12} \\
                 \varrho_{14}^* & -\varrho_{13}^* & \varrho_{12}^* & \frac{1}{2} -\varrho_{11} \\
               \end{array}
             \right).
\label{Eq: second order unpol matrix}\eeq
Here, $\varrho_{11}$ is real and subject to the
restriction $1/6\leq \varrho_{11} \leq 1/3$, whereas $\varrho_{12}$, $\varrho_{13}$, and $\varrho_{14}$ may
be complex. Of course, the general coherence property for the
off-diagonal coefficients $|\varrho_{jk}| \leq \sqrt{\varrho_{jj}\varrho_{kk}}$
holds and imposes additional (but simple) restrictions on the
matrix, once one has chosen $\varrho_{11}$.

Note that the matrix above defines the sufficient conditions for a
density matrix to have vanishing first and second-order
central moments, but that it does not include all necessary conditions
to make it a density matrix. That is, it may be that, for certain
choices of parameters, the matrix is not strictly non-negative. Hence,
the reader is warned that when using Eq.~(\ref{Eq: second order unpol matrix}), to make
sure that the ensuing matrix is non-negative.

For $\oprho$ of the form in Eq.~(\ref{Eq: second order unpol matrix}) one finds that $\langle \hat{\mathbf{S}} \rangle_3 = (0,0,0)$, that $\langle \fluct{i}^2 \rangle_3 = 5$ for $i
= 1,2,3$, and hence that
\beq
\bm{\Gamma}_3 = \left(
                                      \begin{array}{cccc}
                                        5 & 0 & 0 \\
                                        0 & 5 & 0 \\
                                        0 & 0 & 5 \\
                                      \end{array}
                                    \right) .
\eeq

One can also deduce from (\ref{Eq: second order unpol matrix}) that there is no pure, three-photon state that is unpolarized to second order. This follows from the condition that for a pure state, $|\varrho_{jk}|^2=\varrho_{jj}\varrho_{kk}$. Applied to (\ref{Eq: second order unpol matrix}) one gets the three conditions $|\varrho_{12}|^2 = \varrho_{11}(1-3\varrho_{11})$,
$3|\varrho_{12}|^2 = (1-3\varrho_{11})(3 \varrho_{11}-\frac{1}{2})$, and
$|\varrho_{12}|^2 = (3 \varrho_{11}-\frac{1}{2})(\frac{1}{2} -\varrho_{11})$. The first two of these equations demand that $\varrho_{11}=1/3
\rightarrow \varrho_{12}=0$, but this value does not satisfy the third
equation.

An already discussed class of states are the unpolarized states.
This is the smallest class of three-photon states, because there is only one
such 3-photon state. The state has, of course, isotropic
polarization properties of all orders, but requiring this property
for only the lowest three orders uniquely singles out this state.

\begin{figure}
  \includegraphics[width=0.6\columnwidth]{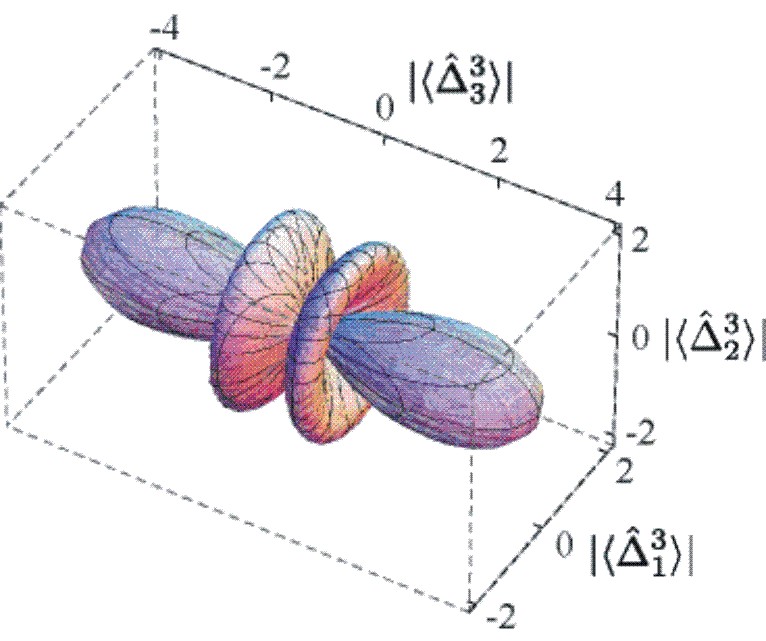}
  \caption{The absolute value of $\langle \fluct{\mathbf{n}}^3 \rangle_3$ for the state $\frac{1}{3}\ket{3,0}\bra{3,0} +
\frac{1}{2}\ket{1,2}\bra{1,2} + \frac{1}{6}\ket{0,3}\bra{0,3}$.}
\label{Fig: State7}
\end{figure}

The mixed state $\frac{1}{3}\ket{3,0}\bra{3,0} +
\frac{1}{2}\ket{1,2}\bra{1,2} + \frac{1}{6}\ket{0,3}\bra{0,3}$ lacks
first-order polarization, has $\langle \fluct{\mathbf{n}}^2 \rangle_3 = 5$, but has third-order polarization structure. Its third-order
polarization central moment is shown in Fig.~\ref{Fig: State7}. This state is thus unpolarized to second order.

The mixed state $(\ket{3,0}\bra{3,0} +  \ket{0,3}\bra{0,3})/2$ has
vanishing first and third-order central moments in all directions, but has an anisotropic second
order polarization central moment, with the predominant fluctuations along
the $\sz$ axis. $\langle \fluct{\mathbf{n}}^2 \rangle_3$ has a ``peanut'' shape (similar to Fig.~\ref{2H,2V_mixture}) with
``semi-axes'' lengths $\langle \fluct{\mathbf{1}}^2 \rangle_3 = \langle \fluct{\mathbf{2}}^2 \rangle_3=3$ and $\langle \fluct{\mathbf{3}}^2 \rangle_3=9$. This is thus a maximum uncertainty state.

\begin{figure}
  \includegraphics[width=0.6\columnwidth]{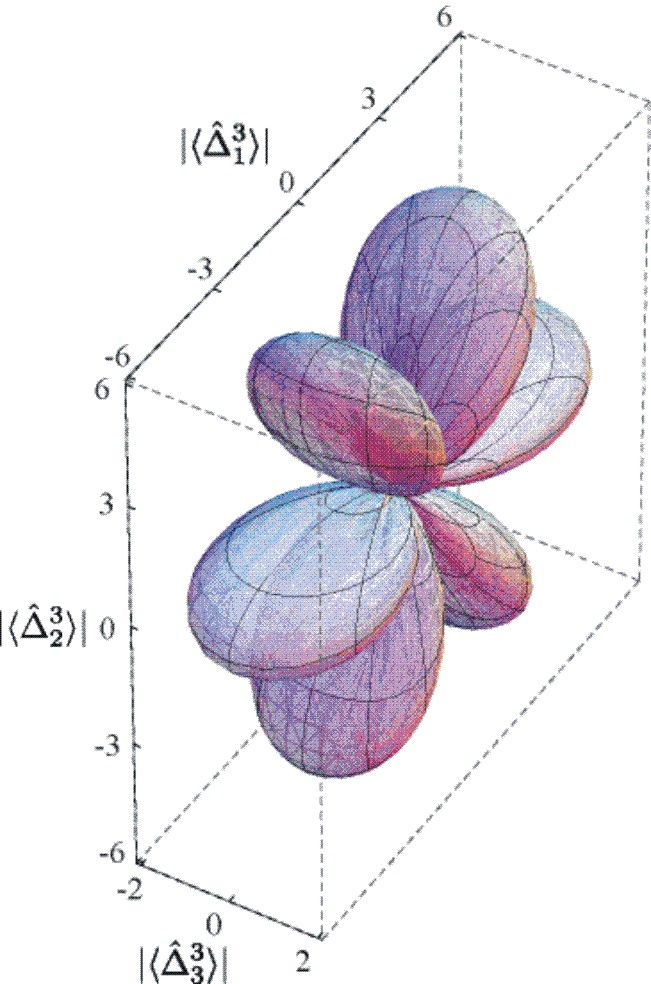}
  \caption{The absolute value of $\langle \fluct{\mathbf{n}}^3 \rangle_3$ for the state $(\ket{0,3} + \ket{3,0})/\sqrt{2}$.}
\label{Fig: State13}
\end{figure}

The pure state $(\ket{0,3} + \ket{3,0})/\sqrt{2}$ lacks first-order polarization, has $\langle \fluct{\mathbf{n}}^2 \rangle_3$ identical to the $(\ket{3,0}\bra{3,0} +  \ket{0,3}\bra{0,3})/2$ mixed state, and its third-order polarization central moment is shown in Fig.~\ref{Fig: State13}. It is also a maximum uncertainty state.

Changing the mixing ratios somewhat, one finds that the mixed state
$\frac{7}{18}\ket{3,0}\bra{3,0} + \frac{1}{3}\ket{1,2}\bra{1,2} +
\frac{5}{18}\ket{0,3}\bra{0,3}$ also has no first-order polarization, but
second- and third-order structure. The second-order
central moment is again ``peanut shaped'' with ``half-axes lengths'' $(19,13,13)/3$, and $\langle \fluct{\mathbf{n}}^3 \rangle_3$ is similar to that in Fig.~\ref{Fig: State7}. Hence, comparing this state with the state $(\ket{0,3} + \ket{3,0})/\sqrt{2}$ they have very similar polarization properties in the first two orders (and both are maximum uncertainty states), but their third-order properties are vastly different.

\begin{figure}
  \includegraphics[width=0.6\columnwidth]{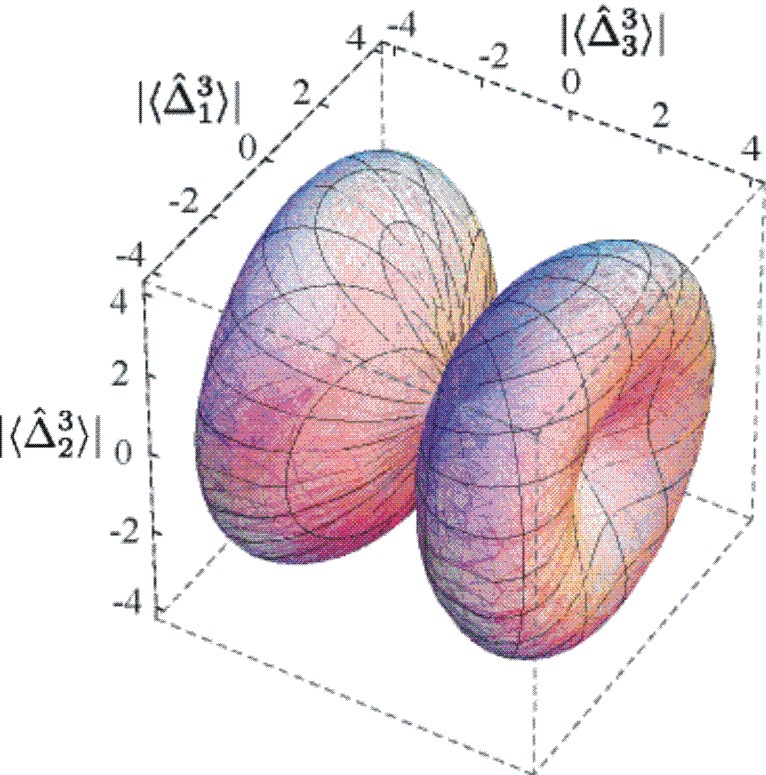}
  \caption{The absolute value of $\langle \fluct{\mathbf{n}}^3 \rangle_3$ for the state $\frac{19}{36}\ket{3,0}\bra{3,0} +
\frac{15}{36}\ket{1,2}\bra{1,2} + \frac{1}{18}\ket{0,3}\bra{0,3}$.}
\label{Fig: State9}
\end{figure}

Finally, an example of a state that has first and third-order
polarization structure but has an isotropic second-order
central moment is the mixed state $\frac{19}{36}\ket{3,0}\bra{3,0} +
\frac{15}{36}\ket{1,2}\bra{1,2} + \frac{1}{18}\ket{0,3}\bra{0,3}$.
This state has the Stokes vector $(-1,0,0)$, $\langle \fluct{\mathbf{n}}^2 \rangle_3 = 14/3$, and $\langle \fluct{\mathbf{n}}^3 \rangle_3$ is shown in Fig.~\ref{Fig: State9}.

It is not possible to find three-photon states that have first-order
polarization but a vanishing third-order polarization central moment.
The above list of possible polarization classes exhausts all the
combination of polarization structures up to third order and shows
that six different classes exist out of the total of eight
\textit{a priori} possible combinations. The classes, with example of associated states, are tabulated in Table \ref{Table1}.

\begin{table*}[h]
\begin{ruledtabular}
\begin{tabular}{|l|c|c|c|}
  \hline
  % after \\: \hline or \cline{col1-col2} \cline{col3-col4} ...
  & \multicolumn{3}{c|}{Polarization transformation invariant}  \\ \hline
  State & $\langle \sind{{\bf n}} \rangle_N$ & $\langle \sind{{\bf n}}^2 \rangle_N$ & $\langle \sind{{\bf n}}^3 \rangle_N$ \\ \hline
  $\hat{1}/4$ & Yes & Yes & Yes \\ \hline
  $\frac{1}{3}\ket{3,0}\bra{3,0} + \frac{1}{2}\ket{1,2}\bra{1,2} +
\frac{1}{6}\ket{0,3}\bra{0,3}$ & Yes & Yes & No  \\ \hline
  $\frac{1}{2}(\ket{3,0}\bra{3,0} + \ket{0,3}\bra{0,3})$ & Yes & No & Yes  \\ \hline
  $(\ket{3,0} + \ket{0,3})/\sqrt{2}$ & Yes & No & No  \\ \hline
  $\frac{19}{36}\ket{3,0}\bra{3,0} +
\frac{15}{36}\ket{1,2}\bra{1,2} + \frac{1}{18}\ket{0,3}\bra{0,3}$ & No & Yes & No  \\ \hline
  $\ket{3,0}$ & No & No & No  \\ \hline
\end{tabular}
\caption{A table of states exemplifying the six different 3-photon polarization classes.}
\label{Table1}
\end{ruledtabular}
\end{table*}

\section{Conclusions}
\label{Sec:Conclusions}

We have developed a systematic method, using central moments, for assessing the polarization characteristics of quantized fields. The method goes well beyond the ``standard'' method that only considers first-order moments, and that moreover, averages over the excitation manifolds. We have shown that there exist a rich ``zoo' of polarization states, including, e.g., states that are unpolarized up to a given order but that have higher-order structure (so called hidden polarization). However, as expected, for most states the polarization characteristics are dominated by the first and second-order behavior, as higher-order polarization moments always contain ``beating terms'' originating from lower orders. Some states, however, show polarization structure that is dominated by higher-order moments, and examples of such states are given.

The suggested method is not the only way to fully characterize the polarization of quantized fields. In particular two more-or-less equivalent methods are mentioned, namely generalized coherence matrices~\cite{klyshko97,Schilling} and the expectation values of all combination of Stokes operators~\cite{soderholm}.

Since a state is not fully specified by its polarization properties, it comes as no surprise that polarization tomography is less resource demanding than full state tomography. We have quantified this difference and indicated a ``recipe'' for determining polarization properties up to a certain order.

\section*{Appendix}

The experiments were performed by using spatially non-degenerate, photon-pair states generated in the process of spontaneous parametric down-conversion. The photon pair centered at 390 nm was generated in a 2 mm thick type-I $\beta$-barium-borate (BBO) crystal pumped by a femtosecond laser pulse centered at 780 nm wavelength. The photon-pairs were subsequently filtered by an interference filter with a 4 nm full width at half maximum (FWHM) bandwidth. The photon pair were brought to the inputs of Hong-Ou-Mandel (HOM)  interferometer \cite{Hong}. When the photons' wavefunction overlap in the HOM interferometer, either the state $\ket{1,1}$ or the state $\ket{2,0}$ can be postselected, dependent on the relative polarizations of the incident photons. The generation setup is described in more detail in \cite{Kwon}.

In order to measure the first and second order Stokes parameters, a polarizing beam splitter (PBS) is positioned at the measurement stage. The measurement basis is changed by means of one half- and one quarter-wave plate which are set in front of the PBS. At each output of the PBS, a two-photon detector is simulated by a 50:50 fiber beam splitter (FB) and two single-photon detectors (PerkinElmer, SPCM-AQRH). The relative coincidence detection-efficiencies are estimated from the FB transmittance. Subsequently, the photon detection-efficiency of each single-photon detector-channel is used to calibrate the measurement of the Stokes parameters. The relative coincidence detection efficiencies of the four detectors are $0.91:0.91:0.82:1$, for $|1,1\rangle$ and $0.75:0.76:1:0.57, \textrm{ for } |2,0\rangle$. In order to achieve full information about the first and second order Stokes parameters, we measured these coincidences in six distinct measurement bases. For a precise measurement, we measured coincidences three times and each measurement is done for 3 s. The central moments are obtained from the measured Stokes operator at the six different directions, and then solving the equation system generated by Eq.~(\ref{Eq: Expansion}).  In the tables, averages of the measurement and the estimated errors due to fluctuations are presented. As can be seen, what looks like errors far exceeding the error bars (e.g., for $\langle \sy \sz + \sz \sy \rangle$ of state $\ket{2,0}$) are actually systematic errors due to imperfect polarization optics. These can be corrected by properly aligning the experimental Poincar\'{e} axes with the ``theoretical'' axes. When this is done, as is shown in Figs. \ref{2Hcut} and \ref{Cutmixture}, the experimental figures are coinciding to within a few percent with the theoretical ones, showing that the errors due to fluctuations are relatively modest and within the relative range indicated by the error limits in the tables.
\begingroup
\squeezetable
\begin{table}[h]
\begin{ruledtabular}
\begin{tabular}{|c|c|c|c|c|}
  \hline
  % after \\: \hline or \cline{col1-col2} \cline{col3-col4} ...
   & \multicolumn{2}{c|}{Theory} & \multicolumn{2}{c|}{Experiment} \\ \hline
  Operator & 1st order & 2nd order & 1st order & 2nd order \\ \hline
  $\langle \sx \rangle$ & 0 & 2 & $-0.19 \pm 0.06$ & $2.06 \pm 0.03$ \\ \hline
  $\langle \sy \rangle$ & 0 & 2 & $0.12 \pm 0.04$ & $2.04 \pm 0.02$ \\ \hline
  $\langle \sz \rangle$ & 2 & 4 & $1.97 \pm 0.01$ & $3.93 \pm 0.02$ \\ \hline
  $\langle \sx \sy + \sy \sx \rangle$ & - & 0 & - & $-0.09 \pm 0.10$\\ \hline
  $\langle \sy \sz + \sz \sy \rangle$ & - & 0 & - & $0.54 \pm 0.08$ \\ \hline
  $\langle \sz \sx + \sx \sz \rangle$ & - & 0 & - & $-0.13 \pm 0.15$ \\ \hline
\end{tabular}
\caption{The experimental data for the state $\ket{2,0}$.}
\label{Table20}
\end{ruledtabular}
\end{table}
\endgroup

\begingroup
\squeezetable
\begin{table}[h]
\begin{ruledtabular}
\begin{tabular}{|c|c|c|c|c|}
  \hline
  % after \\: \hline or \cline{col1-col2} \cline{col3-col4} ...
   & \multicolumn{2}{c|}{Theory} & \multicolumn{2}{c|}{Experiment} \\ \hline
  Operator & 1st order & 2nd order & 1st order & 2nd order \\ \hline
  $\langle \sx \rangle$ & 0 & 4 & $-0.01 \pm 0.04$ & $3.98 \pm 0.00$ \\ \hline
  $\langle \sy \rangle$ & 0 & 4 & $-0.08 \pm 0.03$ & $3.93 \pm 0.03$ \\ \hline
  $\langle \sz \rangle$ & 0 & 0 & $0.01 \pm 0.02$ & $0.15 \pm 0.05$ \\ \hline
  $\langle \sx \sy + \sy \sx \rangle$ & - & 0 & - & $0.03 \pm 0.02$ \\ \hline
  $\langle \sy \sz + \sz \sy \rangle$ & - & 0 & - & $-1.67 \pm 0.14$ \\ \hline
  $\langle \sz \sx + \sx \sz \rangle$ & - & 0 & - & $0.12 \pm 0.16$ \\ \hline
\end{tabular}
\caption{The experimental data for the state $\ket{1,1}$.}
\end{ruledtabular}
\end{table}
\endgroup

\begingroup
\squeezetable
\begin{table}[h]
\begin{ruledtabular}
\begin{tabular}{|c|c|c|c|c|}
  \hline
  % after \\: \hline or \cline{col1-col2} \cline{col3-col4} ...
   & \multicolumn{2}{c|}{Theory} & \multicolumn{2}{c|}{Experiment} \\ \hline
  Operator & 1st order & 2nd order & 1st order & 2nd order \\ \hline
  $\langle \sx \rangle$ & 0 & 8/3 & $-0.07 \pm 0.03$ & $2.69 \pm 0.03$ \\ \hline
  $\langle \sy \rangle$ & 0 & 8/3 & $-0.10 \pm 0.02$ & $2.68 \pm 0.03$ \\ \hline
  $\langle \sz \rangle$ & 0 & 8/3 & $0.01 \pm 0.01$ & $2.67 \pm 0.02$ \\ \hline
  $\langle \sx \sy + \sy \sx \rangle$ & - & 0 & - & $0.00 \pm 0.07$ \\ \hline
  $\langle \sy \sz + \sz \sy \rangle$ & - & 0 & - & $-0.09 \pm 0.06$ \\ \hline
  $\langle \sz \sx + \sx \sz \rangle$ & - & 0 & - & $0.04 \pm 0.09$ \\ \hline
\end{tabular}
\caption{The experimental data for the state $(\ket{2,0}\bra{2,0} + \ket{1,1}\bra{1,1} + \ket{0,2}\bra{0,2})/3$.}
\label{Table:Unpolarized}
\end{ruledtabular}
\end{table}
\endgroup

\begingroup
\squeezetable
\begin{table}[h]
\begin{ruledtabular}
\begin{tabular}{|c|c|c|c|c|}
  \hline
  % after \\: \hline or \cline{col1-col2} \cline{col3-col4} ...
   & \multicolumn{2}{c|}{Theory} & \multicolumn{2}{c|}{Experiment} \\ \hline
  Operator & 1st order & 2nd order & 1st order & 2nd order \\ \hline
  $\langle \sx \rangle$ & 0 & 2 & $-0.11 \pm 0.04$ & $2.04 \pm 0.04$ \\ \hline
  $\langle \sy \rangle$ & 0 & 2 & $-0.10 \pm 0.03$ & $2.05 \pm 0.04$ \\ \hline
  $\langle \sz \rangle$ & 0 & 4 & $0.00 \pm 0.01$ & $3.93 \pm 0.02$ \\ \hline
  $\langle \sx \sy + \sy \sx \rangle$ & - & 0 & - & $-0.01 \pm 0.11$ \\ \hline
  $\langle \sy \sz + \sz \sy \rangle$ & - & 0 & - & $0.69 \pm 0.07$ \\ \hline
  $\langle \sz \sx + \sx \sz \rangle$ & - & 0 & - & $-0.01 \pm 0.11$ \\ \hline
\end{tabular}
\caption{The experimental data for the state $(\ket{2,0}\bra{2,0} +  \ket{0,2}\bra{0,2})/2$.}
\end{ruledtabular}
\end{table}
\endgroup

\begin{acknowledgements}
We thank Prof. G. Leuchs  for useful discussions. Financial support
from  the Swedish Foundation for International Cooperation in Research
and Higher Education (STINT), the Swedish Research Council (VR) through its Linn\ae us Center of Excellence ADOPT and contract 319-2010-7332, the National Research Foundation of Korea (2009-0070668 and 2011-0021452), Spanish DGI (Grants  FIS2008-04356 and FIS2011-26786), the UCM-BSCH program (Grant GR-920992), and
the CONACyT (Grant  106525).
\end{acknowledgements}

\end{document}